\documentclass[aps,prl,twocolumn,showpacs,groupedaddress]{revtex4}
\usepackage{epsfig}
\usepackage{longtable}

\begin{document}
\draft

\title{Test of the isotropy of the speed of light using a continuously rotating optical resonator}
\author{Sven Herrmann$^{1}$, Alexander Senger$^{1}$, Evgeny Kovalchuk$^{1,2}$, Holger M\"{u}ller$^{2,3}$, and Achim Peters$^1$}

\affiliation{$^1$Institut f\"{u}r Physik, Humboldt-Universit\"{a}t
zu Berlin, Hausvogteiplatz 5-7, 10117 Berlin, Germany,}
\email{sven.herrmann@physik.hu-berlin.de,
achim.peters@physik.hu-berlin.de}
\homepage{http://qom.physik.hu-berlin.de/} \affiliation{$^2$
Fritz-Haber-Institut der Max-Planck-Gesellschaft, Faradayweg 4-6,
14195 Berlin, Germany } \affiliation{$^3$ Physics Department,
Stanford University, Stanford, CA 94305, USA}

\date{\today}

\begin{abstract}
We report on a test of Lorentz invariance performed by comparing
the resonance frequencies of one stationary optical resonator and
one continuously rotating on a precision air bearing turntable.
Special attention is paid to the control of rotation induced
systematic effects. Within the photon sector of the Standard Model
Extension, we obtain improved limits on combinations of 8
parameters at a level of a few parts in $10^{-16}$. For the
previously least well known parameter we find $\tilde
\kappa_{e-}^{ZZ} =(-1.9 \pm 5.2)\times 10^{-15}$. Within the
Robertson-Mansouri-Sexl test theory, our measurement restricts the
isotropy violation parameter $\beta -\delta -\frac 12$ to
$(-2.1\pm 1.9)\times 10^{-10}$, corresponding to an eightfold
improvement with respect to previous non-rotating measurements.
\end{abstract}


\maketitle

Local Lorentz invariance (LLI) is an essential ingredient of both
the standard model of particle physics and the theory of general
relativity. It states that locally physical laws are identical in
all inertial reference frames i.e. independent of velocity and
orientation. However, several attempts to formulate a unifying
theory of quantum gravity discuss tiny violations of LLI. Modern
high precision test experiments for LLI are considered as
important contributions to these attempts, as they might either
rule out or possibly reveal the presence of such effects at some
level of measurement precision. An experiment of particular
sensitivity to LLI-violation is the Michelson-Morley (MM)
experiment \cite{MM} testing the isotropy of the speed of light.
Modern versions employ high finesse electromagnetic resonators,
whose eigenfrequencies depend on the speed of light $c$ in a
geometry dependent way ($\nu \sim c/L$ for a linear optical
Fabry-Perot cavity of length $L$). Thus a measurement of the
eigenfrequency of a resonator  as its
orientation is varied, should reveal an anisotropy of $c/L$.\\
Recently, such an anisotropy of $c$ has been described as a
consequence of broken Lorentz symmetry within a test model called
Standard Model Extension (SME) \cite{Kost02}. This model adds all
LLI violating terms that can be formed from the known fields and
Lorentz tensors to the Lagrangian of each sector of the standard
model of particle physics. It thus allows a consistent and
comparative analysis of various experimental tests, including the
MM experiment. The latter however, is also often interpreted
according to a kinematical test theory, formulated by Robertson
\cite{Robertson} and Mansouri and Sexl \cite{MS} (RMS). This test
theory assumes a preferred frame, commonly adopted to be the
cosmic microwave background (CMB). Combinations of three test
parameters ($\alpha$, $\beta$, $\delta$) then model an anisotropy
as
well as a boost dependence of $c$ within a frame moving at velocity $v$ relative to the CMB. \\
In view of the substantial impact that LLI-violation would have
all over physics, the new approach of the SME has triggered a new
generation of improved MM-type experiments
\cite{MMPRL,Wolf,WolfPRD,Lipa}. So far all of these recent
measurements relied solely on Earth's rotation for varying
resonator orientation, which was made possible by the low drift
properties of cryogenically cooled resonators. However, actively
rotating the setup as done in a classic experiment by Brillet and
Hall \cite{BrilletHall} offers two strong benefits: (i) the
rotation rate can be matched to the timescale of optimal resonator
frequency stability and (ii) the statistics can be significantly
improved by performing thousands of rotations per day. While
otherwise using equipment similar to that in the non rotating
experiments, these advantages should allow for tests improved by
orders of magnitude -- assuming that systematic effects induced by
the active rotation can be kept sufficiently low.

\begin{figure}
\centering \epsfig{file=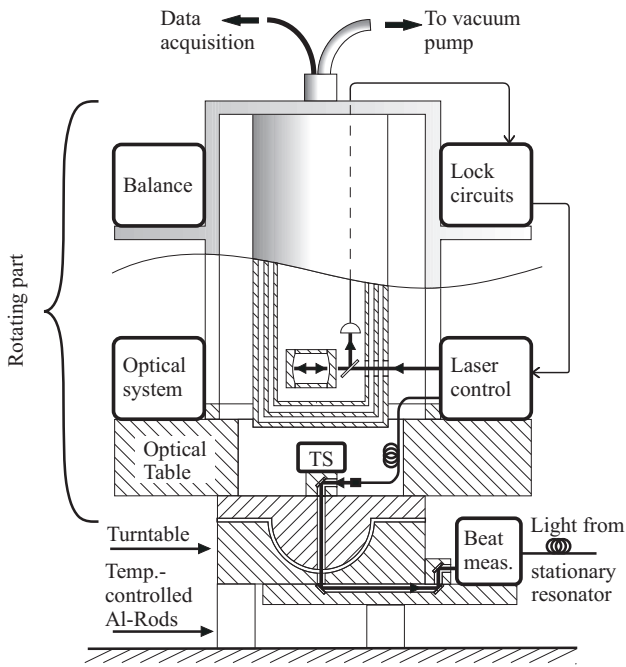, width=0.46\textwidth}
\caption{Setup of the rotating part of the experiment. A high
performance turntable is applied specified for rotation axis
wobble $< 1\mu$rad. The center-of-mass of the setup is carefully
balanced and tilt is monitored using an electronic bubble level
tilt sensor (TS). \label{Setup}}
\end{figure}

\begin{figure}
\centering \epsfig{file=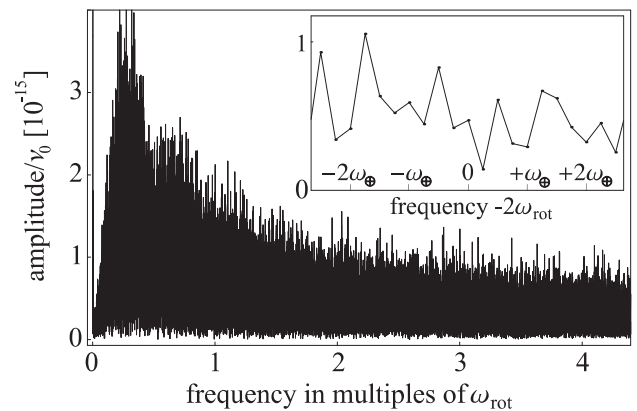, width=0.46\textwidth}
\caption{Fourier transform of a 4-day data set starting on
February 18, 2005, with active tilt control applied (after removal
of long term drift). Insert: No peak is visible at $2\omega_{\rm
rot}$ nor at the siderial sidebands.
 \label{systematics}}
\end{figure}

Here we present the first implementation  of such a continuously
rotating optical MM-type experiment since \cite{BrilletHall}.
Concurrent work of other groups, however, also features similar
experiments either using continuously rotating microwave cavities
\cite{Stanwix} or cryogenic optical resonators, whose orientation
is periodically changed by $90^{\circ}$ \cite{Schiller}. At the
core of the experimental setup is an optical cavity fabricated
from fused silica (L = 3\,cm, 20\,kHz linewidth) which is
continuously rotated on a precision air bearing turntable. Its
frequency is compared to that of a stationary cavity oriented
north-south (L = 10\,cm, 10\,kHz linewidth). Each cavity is
mounted inside a thermally shielded vacuum chamber. The cavity
resonance frequencies are interrogated by two diode pumped Nd:YAG
lasers (1064\,nm), coupled to the cavities through windows in the
vacuum chambers, and stabilized to cavity eigenfrequencies using
the Pound-Drever-Hall method \cite{Dre83}. The table rotation rate
$\omega_{\rm rot} = 2\pi/T$ is set to $T$\,$\sim $\,43\,s
($\sim$\,2000 rotations/day) matching the time scale of optimum
cavity stability ($\Delta \nu / \nu = 1\times 10^{-14}$). At this
rotation rate it is also possible to rely on the excellent thermal
isolation properties of the vacuum chambers at room temperature
(time constant $\sim 10$\,h). The residual temperature drift of
the resonance frequencies is on the order of 1\,MHz/day, which is
comparatively high but sufficiently linear to be cleanly separated
from a potential LLI-violation signal at $2 \omega_{\rm{rot}}$. \\
Fig.\ref{Setup} gives a schematic view of the rotating setup.
Electrical connections are made via an electric 15 contact slip
ring assembly on top. To measure the frequency difference $\Delta
\nu$ of both lasers, a fraction of the rotating laser's light
leaves the table aligned with the rotation axis (see
Fig.\ref{Setup}) and is then overlapped with light from the
stationary laser on a high speed photodetector. The resulting beat
note at the difference frequency $\Delta \nu \sim$2\,GHz is
read out at a sampling rate of 1/s after down conversion to about 100\,MHz. \\
We expended substantial effort on minimizing systematic effects
associated with turntable rotation (see Fig.\ref{systematics}). In
addition to good thermal and electromagnetic shielding, this most
importantly involves limiting cavity deformations due to external
forces (gravitational and centrifugal). If the cavity is not
supported in a perfectly symmetric manner, its frequency is
particularly susceptible to tilt. We observe a relative frequency
change of $1.5 \times 10^{-16}/\mu$rad. As tilts which vary as a
function of the orientation of the turntable enter the analysis of
the experiment, such changes have to be suppressed by keeping the
rotation axis as vertical as possible and preventing wobble in the
setup. The latter is achieved by employing a turntable with
intrinsic wobble $< 1$\,$\mu$rad and carefully balancing the
center-of-mass of the rotating part. To prevent long term
variations of rotation axis tilt, an active tilt control is
applied. Similar to the scheme described by J. Gundlach
\cite{Gundlach}, we place the table on three aluminum cylinders,
$20$\,cm in length, two of which can be heated independently in
order to use thermal expansion ($5\mu$m/$^{\circ}C$) to compensate
slow tilt variations. The heating is part of a computer controlled
closed servo loop, and the tilt is monitored using an electronic
bubble level sensor of 0.1 $\mu$rad resolution placed at the
turntable center. Typical tilt variations of the laboratory's
ground floor are several 10 $\mu$rad/day. Without tilt control
these would give rise to (varying) systematic effects at
$2\omega_{\rm{rot}}$ of up to one part in $10^{-14}$. The active
stabilization reduces tilt variations  to  $< 1 \mu$rad
corresponding to systematic tilt induced effects $ < 10^{-16}$.

\begin{figure}
\centering \epsfig{file=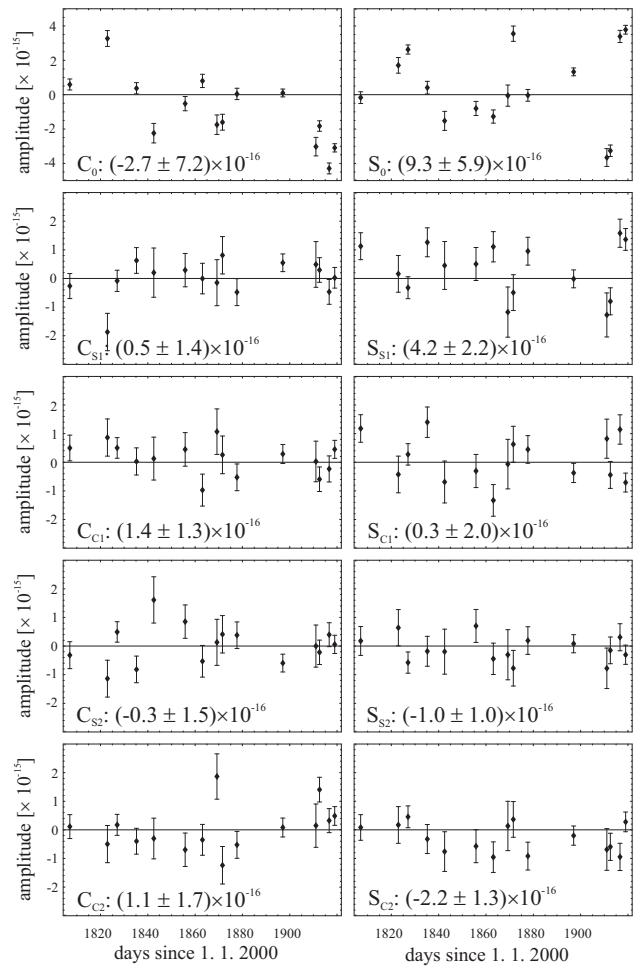, width=0.46 \textwidth}
\caption{Each graph gives the distribution of  a certain Fourier
coefficient of Eq.(\ref{signal2}) and Eq.(\ref{signal3}) in time.
Note the different scale for $C_0$ and $S_0$ affected by small
systematic effects. The time axis spans December 2004 to April
2005. 15 points are included in total, each point is determined
from one continuous data set comprising 2000 - 10000 rotations.
Within each graph the weighted average value of the respective
coefficient is given. \label{data}}
\end{figure}

For our setup, the fundamental signal indicating an anisotropy due
to a LLI-violation is a sinusoidal variation of the beat frequency
at $2\omega_{\rm{rot}}$. As described in \cite{Kost02} the
amplitude of this signal in turn is expected to be modulated due
to Earth's rotation at $\omega_{\oplus}$ (and at Earth's orbital
motion $\Omega_{\oplus}$, which will be considered below). This
can be expressed as
\begin{equation}
\frac{\Delta \nu}{\nu_0} = S(t) \sin2\omega_{\rm rot} t + C(t)
\cos2\omega_{\rm rot} t , \label{signal1}
\end{equation}
where $\nu_0 \sim 2.82 \times 10^{14}$\,Hz is the undisturbed
laser frequency and the amplitudes $S(t)$ and $C(t)$ vary
according to
\begin{eqnarray}
S(t) & = & S_0 + S_{s1} \sin\omega_{\oplus}t + S_{c1}
\cos\omega_{\oplus}t + \label{signal2}
\\ &  & S_{s2} \sin2\omega_{\oplus}t +  S_{c2} \cos2\omega_{\oplus}t , \nonumber \\
C(t) & = & C_0 + C_{s1} \sin\omega_{\oplus}t + C_{c1}
\cos\omega_{\oplus}t
+ \label{signal3} \\
& & C_{s2} \sin2\omega_{\oplus}t +  C_{c2} \cos2\omega_{\oplus}t .
\nonumber
\end{eqnarray}
From each continuous measurement of $\Delta \nu$ comprising 2000
to 10000 rotations, we determine the set of ten Fourier
coefficients $\{S_i,C_i\}$ within Eq.(\ref{signal2}) and
Eq.(\ref{signal3}) in a similar way as in \cite{Schiller}. To
minimize cross-contamination between Fourier coefficients we only
consider data windows that are integer multiples of 24 hours in
length. This method was carefully validated by analyzing test data
sets created by superimposing a hypothetical violation signal to
our data, and checking that the known Fourier coefficients were
reliably reproduced. The procedure is as follows: We divide the
data into subsets of 10 table rotations each (200 subsets/24\,h)
and use a least squares fit to Eq.(\ref{signal1}) for each subset
\cite{phase}. To obtain a proper fit in the presence of drift and
small residual systematics at $\omega_{\rm{rot}}$,  we include
additional sine and cosine components at $\omega_{\rm{rot}}$, an
offset, and a linear and quadratic drift. At the chosen subset
size this is sufficient to cleanly separate the frequency drift
from the signal at $2\omega_{\rm{rot}}$. Next, we fit the
resulting distributions of $S(t)$ and $C(t)$ with
Eq.(\ref{signal2}) and Eq.(\ref{signal3}) \cite{phase} yielding
the complete set $\{S_i,C_i\}$ and individual fit errors for each
coefficient.
\\ Following this scheme we analyzed 15 data sets of 24\,h to
100\,h in length, spanning December 2004 to April 2005 and
comprising $\sim 70000$ turntable rotations in total.
Fig.\ref{data} shows the resulting Fourier coefficients
$\{S_i,C_i\}$ as a function of time together with their weighted
average values. Note that a small systematic effect at
$2\omega_{\rm rot}$ is still present affecting the components
$C_0$ and $S_0$ in particular. This has to be specially considered
within the interpretation of these results according to the two
test theories SME and RMS given below.

\begin{table*}
\caption{Left column: Fourier components $C_i$ related to the SME
parameters for short measurements $\ll 1$ year. These relations
are obtained according to the calculation in \cite{Kost02} and
adopting $\chi = 37.5^{\circ}$ as the laboratory colatitude. $\phi
= \Omega_{\oplus} t$ is the siderial phase relative to t = 0 when
the Earth passes vernal equinox. \\
Right column: $C_i$ related to the RMS-parameter B. $v$ is the
velocity of the laboratory relative to the CMB (neglecting Earth's
orbital and rotational boost here). $\alpha = 168^{\circ}$ and
$\gamma = -6^{\circ}$ fix the orientation of v in the sun centered
reference frame. The respective $S_i$ amplitudes are related
according to $S_0 = 0$, $S_{s1} = -C_{c1}/\cos\chi$, $S_{c1} =
C_{s1}/\cos\chi$, $S_{s2} = -2C_{c2}\cos\chi/(1+\cos^2\chi)$,
$S_{c2} = 2C_{s2}\cos\chi/(1+\cos^2\chi)$.
\label{SME-coefficients}}
\begin{center}
\begin{tabular}{c c c}
\hline \hline  &  SME & RMS \\
\hline $C_0$ & $0.14\tilde\kappa_{e-}^{ZZ}  -   7.4 \times 10^{-6}\tilde \kappa_{o+}^{XY}\cos\phi - 8.5 \times 10^{-6}\tilde \kappa_{o+}^{XZ}\cos\phi  - 9.3 \times 10^{-6}\tilde \kappa_{o+}^{YZ}\sin\phi$ & $\frac{1}{8}(-1 + 3 \cos2\gamma)\sin^2\chi \frac{v^2}{c^2}B $ \\
$C_{s1}$  & $-0.24\tilde \kappa_{e-}^{YZ} + 2.2 \times 10^{-5} \tilde \kappa_{o+}^{XY}\cos\phi -  9.6 \times 10^{-6} \tilde \kappa_{o+}^{XZ}\cos\phi$ &  $-\frac{1}{4}\sin\alpha\sin2\gamma\sin2\chi \frac{v^2}{c^2}B$ \\
$C_{c1}$ & $-0.24\tilde \kappa_{e-}^{XZ}  - 2.4 \times 10^{-5}\tilde \kappa_{o+}^{XY}\sin\phi + 9.6 \times 10^{-6}\tilde \kappa_{o+}^{YZ}\cos\phi$ & $-\frac{1}{4}\cos\alpha\sin2\gamma\sin2\chi \frac{v^2}{c^2}B$ \\
$C_{s2}$ & $0.41\tilde \kappa_{e-}^{XY}  - 4.1 \times 10^{-5}\tilde \kappa_{o+}^{XZ}\sin\phi  - 3.7 \times 10^{-5} \tilde \kappa_{o+}^{YZ}\cos\phi$  & $-\frac{1}{4}\sin2\alpha\cos^2\gamma(1 + \cos^2\chi) \frac{v^2}{c^2}B$\\
$C_{c2}$ & $0.2[\tilde \kappa_{e-}^{XX} - \tilde \kappa_{e-}^{YY}] - 3.7 \times 10^{-5} \tilde \kappa_{o+}^{XZ}\cos\phi + 4.1 \times 10^{-5} \tilde \kappa_{o+}^{YZ}\sin\phi$ & $-\frac{1}{4}\cos2\alpha\cos^2\gamma(1 + \cos^2\chi) \frac{v^2}{c^2}B$ \\
\hline \hline
\end{tabular}
\end{center}
\end{table*}

For the photonic sector of the SME the LLI violating extension
contains 19 independent parameters, which can be arranged into one
scalar $\kappa_{tr}$, and four traceless $3\times 3$ matrices:
$\tilde \kappa_{e-}$, $\tilde \kappa_{o+}$, $\tilde \kappa_{e+}$
and $\tilde \kappa_{o-}$. While $\kappa_{tr}$ is related to the
one way speed of light \cite{Tobar}, the elements of the latter
two matrices are restricted to values $<10^{-32}$ by astrophysical
observations \cite{Kost01}. The remaining matrices $\tilde
\kappa_{e-}$ and $\tilde \kappa_{o+}$ contain 8 parameters that
describe a boost dependent ($\tilde \kappa_{o+}$, antisymmetric)
and a boost independent ($\tilde \kappa_{e-}$, symmetric)
anisotropy of the speed of light. Recent measurements have
restricted 7 of these elements to a level of $10^{-11}$
respectively $10^{-15}$ \cite{MMPRL,Wolf,WolfPRD,Lipa}. $\tilde
\kappa_{e-}^{ZZ}$ can only be determined in actively rotating
experiments thus it was not accessible in these experiments, as
they relied solely on Earth's rotation.  \\
The dependence of the determined Fourier coefficients on these SME
parameters, referred to a Sun centered coordinate system, can be
calculated as outlined in \cite{Kost02}. To first order in orbital
boosts we obtain the combinations given in
Tab.\ref{SME-coefficients}. The amplitudes contain siderial phase
factors, that account for a modulation of the boost dependent
$\tilde \kappa_{o+}$ terms due to Earth's orbit. For data sets
spanning $>1$\, year this allows the independent determination of
$\kappa_{o+}$ and $\kappa_{e-}$ terms by fitting these variations
to the respective distributions of coefficients $\{S_i,C_i\}$.
However, as our data currently only spans 4 months, we can only
extract limits on individual parameters if we additionally assume
no cancellation between the $\tilde \kappa_{e-}$ terms and $\tilde
\kappa_{o+}$ terms. Based on this assumption we obtain the values
given in Tab. \ref{SMElimits}. These limits on the order of few
parts in $10^{-16}$ improve the ones obtained in \cite{WolfPRD} by
up to a factor of eight. A future analysis including data covering
a longer time period will be able to remove the assumption of non-cancellation. \\
\begin{table}
\caption{SME parameters extracted from a fit of the relations of
Tab.\ref{SME-coefficients} to the respective distributions of
Fourier components $\{S_i,C_i\}$ as shown in Fig.\ref{data}. Note
that these limits are based on the assumption of no cancellation
between $\tilde \kappa_{e-}$ and varying $\tilde \kappa_{o+}$
terms. All $\tilde \kappa_{e-}$ values are $\times 10^{-16}$,
$\tilde \kappa_{o+}$ values are $\times 10^{-12}$.
\label{SMElimits}}
\begin{tabular}{c c c c c c}
\hline \hline index & $ZZ$ & $XX - YY$& $XY$ & $XZ$ & $YZ$ \\
\hline  $\tilde \kappa_{e-}$  &  -19.4 (51.8) &  5.4 (4.8) & -3.1 (2.5) & 5.7 (4.9) & -1.5 (4.4)  \\
$\tilde \kappa_{o+}$ & - &- & -2.5 (5.1) & -3.6 (2.7) & 2.9 (2.8) \\
\hline \hline
\end{tabular}
\end{table}
The parameter $\tilde \kappa_{e-}^{ZZ}$ needs a special
consideration as it only enters $C_0$, and might thus be
compromised by the systematic effects. However, we observe that
the phase of this residual systematic signal varies widely between
individual measurements. The systematic effect thus averages out
resulting only in an increased error bar on the mean value of this
component. As the systematic effects are comparatively small, we
can still improve the limit on $\tilde \kappa_{e-}^{ZZ}$ set by
\cite{Stanwix}. From the average value of $C_0$ we deduce a limit
for $(\tilde \kappa_{e-})^{ZZ}$ of $(-1.9 \pm 5.2)\times
10^{-15}$, taking into account that the contributions to $C_0$
from the $\tilde \kappa_{o+}$-terms are already restricted to $<
10^{-15}$ by the other Fourier components. While $(\tilde
\kappa_{e-})^{ZZ}$ plays no special role among the components of
the $\tilde \kappa_{e-}$-matrix, setting such stringent limits on
it is especially important from an experimental point of view, as
it most directly indicates our ability to control rotation related systematic effects.  \\
For comparison to earlier work we also give an analysis within the
RMS framework. This test theory models an anisotropy of the speed
of light according to $\Delta c \sim
B\frac{v^2}{c^2}\sin^{2}\theta$, where $B$ abbreviates the RMS
test parameter combination $(\beta -\delta -\frac 12)$. $v$  is
the laboratory velocity relative to the CMB and $\vartheta$ is the
angle between direction of light propagation and $v$. $B$ enters
the $\{S_i$,$C_i\}$ Fourier amplitudes as shown in
Tab.\ref{SME-coefficients} if we neglect modulation of $v =
370$\,km/s due to orbital boosts. To determine B from our data we
simultaneously fit these functions to the respective distributions
of Fourier coefficients in Fig. \ref{data}, excluding $C_0$
compromised by systematic effects. This results in $B = (-2.1 \pm
1.9)\cdot 10^{-10}$, which is a factor of eight improvement in
accuracy compared to the non rotating experiment of \cite{MMPRL}.

In conclusion, our setup applying precision tilt control proves
that comparatively high rotation rate can be achieved at low
systematic disturbances. This lifts a severe limitation from
actively rotated MM-type experiments as performed in the past
\cite{BrilletHall}, and provides the possibility to increase
sensitivity of these tests to LLI-violation by orders of
magnitude. At the current status of our measurement we can already
set limits on several test theory parameters that are more
stringent by up to a factor of eight. An extended analysis of the
experiment within the SME shows that it is also sensitive to
parameters from the electronic sector of the SME that change the
cavity length \cite{MuePRD}. While this provides the possibility
to set limits on further SME parameters, we leave it for a future
analysis. The main limitation of accuracy within our experimental
setup currently arises from laser lock stability. Thus, the
implementation of an active vibration isolation as well as new
cavities is underway, which should enable
us to improve laser lock stability by about an order of magnitude.\\
We thank Claus L\"{a}mmerzahl for discussions and J\"{u}rgen
Mlynek and Gerhard Ertl for making this experiment possible. S.
Herrmann acknowledges support from the Studienstiftung des
deutschen Volkes.


\begin{references}
\bibitem{phase} The phase of the respective fits is fixed by the
test theory considered according to \cite{Kost02}.
\end{references}
\end{document}